\documentclass{article}
\usepackage{eurosym}
\usepackage{amssymb}

\newtheorem{theorem}{Theorem}

\newtheorem{proposition}{Proposition}

\newtheorem{example}{Example}

\begin{document}

\title{New methods to compensate artists in music streaming platforms\thanks{%
Financial support from grants PID2020-113440GBI00 and PID2023-146364NB-I00,
funded by MCIN/AEI/ 10.13039/501100011033 and
MICIU/AEI/10.13039/501100011033/ respectively, and by FEDER, and UE is
gratefully acknowledged. We also thank participants of seminars at Rabat,
Buenos Aires, Sao Paulo, Santiago and Santander, where earlier versions of this paper
were presented, for helpful comments and suggestions.}}
\author{\textbf{Gustavo Berganti\~{n}os}\thanks{%
ECOBAS, Universidade de Vigo, ECOSOT, 36310 Vigo, Espa\~{n}a} \\
%EndAName
\textbf{Juan D. Moreno-Ternero}\thanks{%
Department of Economics, Universidad Pablo de Olavide, 41013 Sevilla, Espa%
\~{n}a; jdmoreno@upo.es}}
\maketitle

\begin{abstract}
We study the problem of measuring the popularity of artists in music
streaming platforms and the ensuing methods to compensate them (from the
revenues platforms raise by charging users). We uncover the space of
popularity indices upon exploring the implications of several axioms
capturing principles with normative appeal. As a result, we characterize several families of
indices. Some of them are intimately connected to the Shapley value, the central tool in
cooperative game theory. Our characterizations might help to address the rising concern in the music industry to explore new methods that reward artists more appropriately. We actually connect our families to the new royalties models, recently launched by Spotify and Deezer. 

\bigskip

\medskip {} %C72, D21, L83


\bigskip
\end{abstract}

\newpage

\section{Introduction}

The music industry has evolved drastically in recent years. The advent of
streaming, followed by the effect of COVID-19, permitted platforms such as
Spotify (and, to a lower extent, Amazon Music, Deezer, or Apple Music) to
become major actors, with millions of customers worldwide. For instance, in
2011, Spotify announced a customer base of 1 million paying subscribers
across Europe, and was officially launched in the US. In the %fourth quarter of 2024, it reached an all-time high with 675 million active users worldwide, an increase of 12 percent in just one year. 
second quarter of 2025, it reached an all-time high with 696 million active users worldwide, an increase of 70 million in just one year. 
%Spotify continues to dominate the global music streaming industry, boasting 696 million monthly active users, according to the latest data. This is an increase of 18 million MAUs from Q1 2025 (678 million).
The initial concerns about the potential negative effects that streaming might had on the music industry have almost vanished. For instance, Aguiar and Waldfogel (2018) show that streaming displaces music piracy and that the losses from displaced sales are roughly outweighed by the gains in streaming revenue.\footnote{Aguiar et al. (2021) also showed that concerns of platform biases in favor of major record labels seem to be unfounded.} Somewhat related, Christensen (2022) shows that streaming stimulates the live concert industry. Overall, streaming
now accounts for $69\%$ of recorded music trade revenue. Thus, one
might say that the old tradition of buying physical albums (CDs, vinyls, or
cassettes) in record stores is gone, and customers now essentially pay monthly
subscription fees to have on-demand access for the (almost unlimited)
catalogues of music platforms, or access a free (ad-sponsored and with
restricted freedom of choice) version (e.g., Aguiar and Martens, 2016;
Aguiar et al., 2024).
This modern instance of bundling echoes some of the standard features of
bundling that are well known in the industrial organization literature
(e.g., Adams and Yellen, 1976; Shiller and Waldfogel, 2011, 2013;
Belleflamme and Peitz, 2015). In particular, a special effort has to be made
on the ensuing problem of sharing the revenue raised from paid subscriptions
to streaming platforms among artists, as it is estimated that streaming platforms redistribute among ``right holders" (the artists themselves if they are independent, or the record labels if the artist is signed to one) around $65-70\%$ of the revenue they raise.

It is claimed that the revenue distribution problem in streaming has not been sufficiently studied yet (e.g., UNESCO, 2022).\footnote{This is in contrast to what
happens with other revenue sharing problems under bundled pricing (e.g.,
Liao and Tauman, 2002; Ginsburgh and Zang, 2003; Pagnozzi, 2009; Berganti\~{n}os and Moreno-Ternero, 2015; Chen and Ni, 2017).} %From a different vantage point, Etro (2023) analyzes platforms setting access prices and commissions on revenues of sellers engaged in monopolistic competition with free entry.  
Nevertheless, there has been a recent literature dealing with the problem of
revenue sharing in music platforms (e.g., Alaei et al., 2022; Lei, 2023;
Moreau et al., 2024; Schlicher et al., 2024; Berganti\~{n}os and
Moreno-Ternero, 2025a, 2025b, 2026; Gon\c{c}alves-Dosantos et al., 2025a,
2025b). A central impetus in this literature has been to understand the
methods that platforms were using in their early years; mostly, the 
\textit{pro-rata} method, in which artists are rewarded in proportion of
their total streams and the \textit{user-centric} method, in which, instead,
the amount paid by each user is shared among the artists this user streamed,
in proportion of the user's overall streams. Other methods, either
compromising between the above two or extending some of their features, have
been proposed too, and the aim of this paper is to uncover the space of
those methods upon exploring the implications of several axioms capturing
principles with normative appeal. In doing so, we hope to provide solid
foundations for alternative compensation methods, which might contribute to
the ongoing debate in the music industry and the general public alike.%
%\footnote{GUS: Se podr\'{\i}a incluir un resumen m\'{a}s exhaustivo de lo que hacemos. Por ejemplo mencionando los axiomas y las caracterizaciones axiom\'{a}ticas. Esto puede ser interesante dependiendo d ela revista a lo que lo mandemos.}

The rest of the paper is organized as follows. In Section 2, we introduce
the model and the basic definitions (including our axioms). In Section 3, we
provide the characterization theorems for our families of indices. In
Section 4, we connect these families with other existing ones in the
literature. In Section 5, we also connect them to the recently launched methods in the music industry. We conclude in Section 6. For a smooth passage, we defer the
proofs of all results to the Appendix.

\section{Streaming problems}

\subsection{Preliminaries}

We consider the model introduced by Berganti\~{n}os and Moreno-Ternero
(2025a). Let $\mathbb{N}$ represent the set of all potential artists and $%
\mathbb{M}$ the set of all potential users (of music streaming platforms).
We assume that both $\mathbb{N}$ and $\mathbb{M}$ are sufficiently large.
Each specific platform involves a specific (finite) set of artists $N\subset 
\mathbb{N}$ and a specific (finite) set of users $M\subset \mathbb{M}$. We
typically assume $N=\left\{ 1,...,n\right\} $ and $M=\left\{ 1,...,m\right\} 
$. For each $(i,j)\in N\times M$, let $t_{ij}$ denote the times user $j$
streamed contents uploaded by artist $i$ in the platform (briefly, \textit{%
streams}), during a certain period of time. In most of the platforms,
playing a streaming unit will be equivalent to playing a song (for at least
30 seconds). Let $t=\left( t_{ij}\right) _{i\in N,j\in M}$ denote the
corresponding matrix encompassing all streams. We assume that for each $j\in
M,$ $\sum\limits_{i\in N}t_{ij}>0$ (namely, each user has streamed some
content).

A \textbf{streaming problem} is a triple $P=\left( N,M,t\right) $. We
normalize the amount paid by each user in such a way that the amount to be
divided among artists in a problem is $m,$ the number of users. The set of
problems so defined is denoted by $\mathcal{P}$.

%\bigskip

The next example will be used to illustrate some concepts introduced throughout the
paper.

\begin{example}
\label{ex 2,3} Let $N=\left\{ 1,2\right\} ,$ $M=\left\{ a,b,c\right\} $ and 
\[
t=\left( 
\begin{array}{ccc}
100 & 0 & 10 \\ 
0 & 10 & 20%
\end{array}%
\right) 
\]%
where $t_{1a}=100,$ $t_{2a}=0$ and so on.
\end{example}

\bigskip

For each $j\in M,$ we denote by $t^{-j}$ the matrix obtained from $t$ by
removing the column corresponding to user $j$. Similarly, for each $i\in N,$
we denote by $t_{-i}$ the matrix obtained from $t$ by removing the row
corresponding to artist $i$.

For each artist $i\in N$, we denote the total number of streams of artist $i$
as 
\[
T_{i}\left( N,M,t\right) =\sum_{j\in M}t_{ij}. 
\]

For each user $j\in M,$ we denote the total number of streams of user $j$ as 
\[
T^{j}\left( N,M,t\right) =\sum_{i\in N}t_{ij}, 
\]

The set of fans of each artist is defined as the set of users who have
played content from the artist. Formally, for each $i\in N,$ 
\[
F_{i}\left( N,M,t\right) =\left\{ j\in M:t_{ij}>0\right\} . 
\]

Similarly, we define the list of artists of a user as those from which the
user has played content. Formally, for each $j\in M,$ 
\[
L^{j}\left( N,M,t\right) =\left\{ i\in N:t_{ij}>0\right\} . 
\]

The profile of user $j$ is defined as the streaming vector associated to the
user. Namely, 
\[
t_{.j}\left( N,M,t\right) =\left( t_{ij}\right) _{i\in N}. 
\]

When no confusion arises we write $T_{i}$ instead of $T_{i}\left(
N,M,t\right) $, $T^{j}$ instead of $T^{j}\left( N,M,t\right) ,$ $F_{i}$
instead of $F_{i}\left( N,M,t\right) ,$ $L^{j}$ instead of $L^{j}\left(
N,M,t\right) ,$ and $t_{.j}$ instead of $t_{.j}\left( N,M,t\right) .$

\subsection{Indices}

A popularity \textbf{index} $\left( I\right) $ is a mapping that yields the
importance of each artist in each problem. Formally, $I:\mathcal{P}%
\rightarrow \mathbb{R}_{+}^{N}$ is such that for each pair $i,j\in N$, $%
I_{i}\left( N,M,t\right) \geq I_{j}\left( N,M,t\right) $ if and only if $i$
is at least as important as $j$ at problem $\left( N,M,t\right) $. We assume
that $\sum\limits_{i\in N}I_{i}\left( N,M,t\right) >0$.

The \textbf{reward} received by each artist $i\in N$ in each problem is
proportional to the importance of that artist in that problem. Formally, 
\[
R_{i}^{I}\left( N,M,t\right) =\frac{I_{i}\left( N,M,t\right) }{%
\sum\limits_{i^{\prime }\in N}I_{i^{\prime }}\left( N,M,t\right) }m. 
\]

As for each $\lambda >0$ and each index $I$, $R^{\lambda I}=R^{I}$, we
identify an index with all its positive linear transformations.

The index used by most platforms is the so called \textbf{pro-rata} index,
in which the importance is the total number of streams. Formally, for each
problem $\left( N,M,t\right)\in \mathcal{P} $ and each artist $i\in N$,  
\[
P_{i}\left( N,M,t\right) =T_{i}=\sum_{j\in M}t_{ij}. 
\]

Thus, the amount received by each artist $i\in N$ under $P$ is

\[
R_{i}^{P}\left( N,M,t\right) =\frac{T_{i}}{\sum\limits_{j\in N}T_{j}}m. 
\]

%Pro-rata has been studied in Berganti\~{n}os and Moreno-Ternero (2024b, 2025a) and Gon\c{c}alves-Dosantos $et$ $al$ (2025a, 2025b).

Another index that has been increasingly used in the music industry is the
so called \textbf{user-centric} index. In this case, all users are assumed
to be equally important (which is normalized to 1). The importance of each
user is divided among the artists streamed by this user proportionally to
the total number of streams. Then, the importance of each artist is the sum,
over all users, of the importance of the artist given by each user.
Formally, for each problem $\left( N,M,t\right) $ and each artist $i\in N,$ 
\[
U_{i}\left( N,M,t\right) =\sum_{j\in M}\frac{t_{ij}}{T^{j}}. 
\]

It is straightforward to show that the index and the rewards coincide in
this case. Namely, for each $i\in N,$ $R_{i}^{U}\left( N,M,t\right)
=U_{i}\left( N,M,t\right) .$ 
%User-centric has been studied in Berganti\~{n}os and Moreno-Ternero (2024b, 2025a) and Gon\c{c}alves-Dosantos $et$ $al$ (2025a, 2025b).

%\bigskip

%In Example \ref{ex 2,3} we have the following: 
%\[
%\begin{tabular}{cccc}
%& $P_{i}\left( N,M,t\right) $ & $R_{i}^{P}\left( N,M,t\right) $ & $%
%U_{i}\left( N,M,t\right) $ \\ 
%1 & 110 & 2.36 & 1.33 \\ 
%2 & 30 & 0.64 & 1.66%
%\end{tabular}%
%\]

\bigskip

%\begin{remark} \label{remark equal-division}Gon\c{c}alves-Dosantos $et$ $al$ (2025a, 2025b) consider the \textbf{equal-division} index which assigns the same importance to each artist. Formally, for each problem $\left( N,M,t\right) $ and each artist $i\in N,$ $E_{i}\left( N,M,t\right) =\frac{m}{n}.$ It is trivial to argue that the index and the reward coincide. Namely, for each $i\in N,$ $R_{i}^{E}\left( N,M,t\right) =E_{i}\left( N,M,t\right) .$
%\end{remark}

Berganti\~{n}os and Moreno-Ternero (2025a) introduced the following family
of indices offering a natural compromise between the pro-rata and the
user-centric indices. A weight system is a function $\omega :\mathbb{M}%
\times \mathbb{Z}_{+}^{n}\rightarrow \mathbb{R}$ such that for each $j\in 
\mathbb{M}$ and each $x\in \mathbb{Z}_{+}^{n},$ $\omega \left( j,x\right)
>0. $ Given a weight system $\omega $, the \textbf{weighted index} $%
I^{\omega }$ is defined as follows. For each $\left( N,M,t\right) \in 
\mathcal{P}$ and each $i\in N,$ 
\[
I_{i}^{\omega }\left( N,M,t\right) =\sum_{j\in M}\omega \left(
j,t._{j}\right) t_{ij}. 
\]

To conclude this initial inventory of indices, Berganti\~{n}os and
Moreno-Ternero (2025b) consider an index that does not belong to the
previous family. For reasons that will become clear later in the text, this
index is called the \textbf{Shapley} index. It divides the amount paid by
each user among the artists streamed by the user equally (thus, ignoring the
streaming times; as opposed to the user-centric index which did it
proportionally to the streaming times). Formally, for each streaming problem 
$\left( N,M,t\right) $ and each $i\in N,$ 
\[
Sh_{i}\left( N,M,t\right) =\sum_{j\in M:i\in L^{j}}\frac{1}{\left\vert
L^{j}\right\vert }.
\]%
As $\sum\limits_{i\in N}Sh_{i}\left( N,M,t\right) =m$, the index and the
reward associated coincide too. Namely, $Sh\left( N,M,t\right)
=R^{Sh_{i}\left( N,M,t\right) }$. 
%Thus, we define \textbf{Shapley} as $Sh\left( N,M,t\right) $.

In Example \ref{ex 2,3} we have the following: 
\[
\begin{tabular}{cccc}
& $R_{i}^{P}\left( N,M,t\right) $ & $U_{i}\left( N,M,t\right) $ & $%
Sh_{i}\left( N,M,t\right) $ \\ 
1 & 2.36 & 1.33 & 1.5 \\ 
2 & 0.64 & 1.66 & 1.5%
\end{tabular}%
\]

\subsection{Axioms}

We now introduce some axioms of indices, already considered in Berganti\~{n}%
os and Moreno-Ternero (2025a, 2025b) reflecting principles with normative
appeal. 
%We now introduce the axioms used in this paper, which have been already considered in Berganti\~{n}os and Moreno-Ternero (2024a, 2025).
%\bigskip
% The first two axioms: additivity and null artists are quite standard in the literature and will be used in all our results. Both axioms have been considered in Berganti\~{n}os and Moreno-Ternero (2024a, 2025) and Gon\c{c}alves-Dosantos $et$ $al$ (2025a, 2025b). Additivity is called composition in Gon\c{c}alves-Dosantos $et$ $al$ (2025a, 2025b) whereas null artists is called null service in Gon\c{c}alves-Dosantos $et$ $al$ (2025a) and nullity in Gon\c{c}alves-Dosantos $et$ $al$ (2025b),

First, a standard axiom in the literature that can be traced back to Shapley
(1953). It says that if we can divide a problem as the sum of two smaller
problems, then the solution to the original problem should be the sum of the
solutions to the two smaller problems. Formally,

\textbf{Additivity}. For each triple $\left( N,M^{1},t^{1}\right) ,\left(
N,M^{2},t^{2}\right) ,\left( N,M,t\right) \in \mathcal{P}$ such that $%
M=M^{1}\cup M^{2}$, $M^{1}\cap M^{2}=\varnothing ,$ $t_{ij}=t_{ij}^{1}$ when 
$j\in M^{1}$ and $t_{ij}=t_{ij}^{2}$ when $j\in M^{2},$ 
\[
I\left( N,M,t\right) =I\left( N,M^{1},t^{1}\right) +I\left(
N,M^{2},t^{2}\right) . 
\]

%\bigskip

The next axiom is also standard and says that if an artist is not streamed,
then its importance should be zero. Formally,

\textbf{Null artists.} For each $\left( N,M,t\right) \in \mathcal{P}$, and
each $i\in N$ such that $T_{i}=0$, 
\[
I_{i}\left( N,M,t\right) =0. 
\]

%Berganti\~{n}os and Moreno-Ternero (2024a, 2025) argue that both axioms are satisfied by the three index defined above.
%\bigskip

A stronger axiom is \textit{pairwise homogeneity}, which says that if each
user streams an artist certain times more than another artist, the index
should preserve that ratio. 
%Two stronger axioms than the previous one have been considered in the emerging literature on streaming problems. First, \textit{pairwise homogeneity}, which says that if each user streams an artist certain times more than another artist, the index should preserve that ratio. Second, \textit{reasonable lower bound}, which says that given a set of users $C$ and the set of artists $A$ streamed by those users, the amount received by the artists in $A$ should be, at least, the amount paid by users in $C$. 
Formally,

\textbf{Pairwise homogeneity.} For each $\left( N,M,t\right) \in \mathcal{P}$%
, each pair $i,i^{\prime }\in N,$ and each $\lambda \geq 0$ such that $%
t_{ij}=\lambda t_{i^{\prime }j}$ for all $j\in M,$ $\ $ 
\[
I_{i}\left( N,M,t\right) =\lambda I_{i^{\prime }}\left( N,M,t\right) . 
\]

%\textbf{Reasonable lower bound. }%$\mathbf{FUP}$. 
%For each $\left( N,M,t\right) \in \mathcal{P}$ and each $C\subset M$, let $L^{C}=\bigcup\limits_{j\in C}L^{j}.$ Then, 
%\[
%\sum_{i\in L^{C}}\frac{I_{i}\left( N,M,t\right) }{\sum\limits_{k\in N}I_{k}\left( N,M,t\right) }m\geq \left\vert C\right\vert .
%\]

%It is straightforward to see that each of the previous two axioms implies the axiom of null artists. 

To conclude with this inventory of axioms, we introduce three axioms that
say, roughly speaking, when streams, users, or artists (the three
ingredients of streaming problems) should have the same impact on the index.

The first one applies to streams. It 
%, introduced in Berganti\~{n}os and Moreno-Ternero (2025), 
says that the index of each artist depends on the number of streams of users
but not on the user who produce the streams. Namely, if two users have the
same streams on a given artist, then, both users should have the same impact
over this artist. Formally,

\textbf{Equal individual impact of similar users}. For each $\left(
N,M,t\right) \in \mathcal{P}$, each $i\in N,$ and each pair $j,j^{\prime
}\in M$ such that $t_{ij}=t_{ij^{\prime }}$, 
\[
I_{i}\left( N,M\backslash \left\{ j\right\} ,t^{-j}\right) =I_{i}\left(
N,M\backslash \left\{ j^{\prime }\right\} ,t^{-j^{\prime }}\right) . 
\]

%\bigskip

The second one 
%, introduced also in Berganti\~{n}os and Moreno-Ternero (2025), 
applies to users. It says that all users should have the same impact on the
index. Formally,

\textbf{Equal global impact of users. } For each $\left( N,M,t\right) \in 
\mathcal{P}$ and each pair $j,j^{\prime }\in M,$ 
\[
\sum_{i\in N}I_{i}\left( N,M\backslash \left\{ j\right\} ,t^{-j}\right)
=\sum_{i\in N}I_{i}\left( N,M\backslash \left\{ j^{\prime }\right\}
,t^{-j^{\prime }}\right) . 
\]

%\bigskip

The third one %, introduced in Berganti\~{n}os and Moreno-Ternero (2024a),
applies to artists. It says that for any pair of artists $i$ and $i^{\prime
},$ the impact of artist $i$ in the index of artist $i^{\prime }$ is the
same as the impact of artist $i^{\prime }$ in the index of artist $i.$
Namely, if artist $i$ leaves the platform, the change in the index of any
other artist $i^{\prime } $ coincides with the change in the index to artist 
$i$ when artist $i^{\prime }$ leaves the problem. Formally,

\textbf{Equal impact of artists}. For each $\left( N,M,t\right) \in \mathcal{%
P}$ and each pair $i,i^{\prime }\in N$, 
\[
I_{i}\left( N,M,t\right) -I_{i}\left( N\backslash \left\{ i^{\prime
}\right\} ,M,t_{-i^{\prime }}\right) =I_{i^{\prime }}\left( N,M,t\right)
-I_{i^{\prime }}\left( N\backslash \{i\},M,t_{-i}\right) . 
\]

%This axiom is inspired in the ideas of the axiom of balanced contributions for cooperative games (Myerson, 1980).

\section{Results}

We reproduce first a result in Berganti\~{n}os and
Moreno-Ternero (2025a), which will be the starting point of our analysis here. 

\begin{proposition}
\label{MS-th} The following statements hold:

$\left( a\right) $ An index satisfies \textit{additivity} and \textit{\
pairwise homogeneity} if and only if it is a weighted index.

$\left( b\right) $ An index satisfies \textit{additivity}, \textit{pairwise
homogeneity}, and \textit{equal individual impact of similar users} if and
only if it is the pro-rata index.

$\left( c\right) $ An index satisfies \textit{additivity}, \textit{pairwise
homogeneity}, and \textit{equal global impact of users} if and only if it is
the user-centric index.
\end{proposition}

Our first result in this paper (Theorem \ref{char NA+AdU} below) generalizes
Proposition \ref{MS-th} upon replacing pairwise homogeneity by the weaker
axiom of null artist. To present it, we need some notation and definitions
first.

%We obtain several axiomatic characterizations. In the first one we characterize the set of indices satisfying our two basic axioms. additivity and null artists. In such indices the importance of each artist is computed as the sum, over all indices, of the importance of the artist for each user. Later on we add to the two basic axioms each one of the three \textquotedblleft equal impact\textquotedblright\ axioms and we characterize three set of indexes. Finally, we add to the two basic axioms any possible combination of two of the three \textquotedblleft equal impact\textquotedblright\ axioms. In this case no index satisfies the two basic axioms together with equal individual impact of similar users and equal global impact of users. For the other two possible combinations we characterize the associated families of indices. \bigskip

We say that $I$ is a \textbf{decomposable index} if the following two
conditions hold:
\begin{enumerate}
\item There exists a function $d:$ $N\times M\times \mathbb{N}%
_{+}^{N}\rightarrow \mathbb{R}_{+}$ such that for each $i\in N$, $j\in M$,
and $x\in $ $\mathbb{N}_{+}^{N}$, $d\left( i,j,x\right) =0$ whenever $%
x_{i}=0.$

\item For each $\left( N,M,t\right)\in\mathcal{P} $ and each $i\in N$, 
\[
I_{i}\left( N,M,t\right) =\sum_{j\in M}d\left( i,j,t._{j}\right) . 
\]
\end{enumerate}

For each function $d$ satisfying conditions $1$ and $2$, we denote by $I^{d}$
the decomposable index associated with $d$.

%\bigskip

Notice that the three indices defined above (pro-rata, user-centric, and
Shapley) are decomposable indices. For the pro-rata index, consider $d\left(
i,j,x\right) =x_{i}$. For the user-centric index, consider $d\left(
i,j,x\right) =\frac{x_{i}}{\sum\limits_{i^{\prime }\in N}x_{i^{\prime }}}.$
For the Shapley index, consider $d\left( i,j,x\right) =0$ if $x_{i}=0$ and $%
d\left( i,j,x\right) =\frac{1}{\left\vert i^{\prime }\in N:x_{i^{\prime
}}>0\right\vert }$ if $x_{i}>0$. Note also that all weighted indices are
decomposable indices too.

\bigskip

We say that a decomposable index $I^{d}$ is 
%$d:$ $N\times M\times \mathbb{N}_{+}^{N}\rightarrow \mathbb{R}_{+}$ 
\textbf{user independent} if given $\left( i,j,x\right) $ and $\left(
i,j^{\prime },x^{\prime }\right) $ such that $x_{i}=x_{i}^{\prime }$ we have
that $d\left( i,j,x\right) =d\left( i,j^{\prime },x^{\prime }\right).$

Note that pro-rata belongs to this family of indexes, whereas user-centric
and Shapley do not. 
%Berganti\~{n}os and Moreno-Ternero (2025) prove that pro-rata is the unique index satisfying pairwise homogeneity, additivity, and equal individual impact of similar users. Since any index satisfying pairwise homogeneity also satisfies null artists (Berganti\~{n}os and Moreno-Ternero, 2024a), Theorem \ref{char NA+AdU+EIISU} can be seen as a generalization of the previous result.

%\bigskip
%The second axiom we consider is equal global impact of users.

%\bigskip

Finally, we say that a decomposable index $I^{d}$ 
%$d:$ $N\times M\times \mathbb{N}_{+}^{N}\rightarrow \mathbb{R}_{+}$
is \textbf{aggregate invariant} if for all $j,j^{\prime }\in M$ and all $%
x,x^{\prime }\in \mathbb{N}_{+}^{N}$ we have that $\sum\limits_{i\in
N}d_{i}\left( i,j,x\right) =\sum\limits_{i\in N}d_{i}\left( i,j^{\prime
},x^{\prime }\right) .$

Note that the user-centric and Shapley indices belong to the previous
family, whereas pro-rata does not.

%\bigskip
%Berganti\~{n}os and Moreno-Ternero (2025) prove that user-centric is the unique index satisfying pairwise homogeneity, additivity, and equal global impact of users. Since any index satisfying pairwise homogeneity also satisfies null artists, Theorem \ref{char NA+AdU+EIISU} can be seen as a generalization of the previous result. There exist indexes, different from user-centric, that belong to the family of indexes given by Theorem \ref% {char NA+AdU+EIISU}. For instance, the Shapley index.
%\bigskip

%\bigskip
%It turns out, as the next theorem states, that the family of decomposable indices is characterized by the combination of additivity and null artists.

\medskip

We are now ready to present the first original characterizations in this
paper.

\begin{theorem}
\label{char NA+AdU} The following statements hold:

$\left( a\right) $ An index satisfies additivity and null artists if and
only if it is a decomposable index.

$\left( b\right) $ An index satisfies additivity, null artists, and equal
individual impact of similar users if and only if it is a user-independent
decomposable index.

$\left( c\right) $ An index satisfies additivity, null artists, and equal
global impact of users if and only if it is an aggregate-invariant
decomposable index.
\end{theorem}

%In the following results we add the three equal impact axioms to the ones used in Theorem \ref{char NA+AdU} and we obtain new families of indexes. The first axiom we consider is equal individual impact of similar users. \bigskip

%\begin{theorem}
%\label{char NA+AdU+EIISU}An index $I$ satisfies additivity, null artists,  and equal individual impact of similar users if and only if $\ I$ is a user-independent decomposable index.
%\end{theorem}

%\bigskip

%\begin{theorem}
%\label{char NA+AdU+EGIU}An index $I$ satisfies additivity, null artists,  and equal global impact of users if and only if $\ I$ is a aggregate-invariant decomposable index.
%\end{theorem}

Our next result characterizes the family of indices that arises from
combining additivity and null artists with equal impact of artists. It turns
out that this family has strong cooperative game-theory underpinnings, as
explained next.

A \textit{cooperative game with transferable utility}, briefly a \textit{TU
game}, is a pair $\left( N,v\right) $, where $N$ denotes a set of agents and 
$v:2^{N}\rightarrow \mathbb{R}$ satisfies $v\left( \varnothing \right) =0.$
Given a $TU$ game $\left( N,v\right) $ and $S\subset N,$ we define the game $%
\left( S,v\right) $ as the restriction of $v$ to agents in $S.$

Given $N\in \mathbb{N}$, let $\Pi _{N}$ denote the set of all orders on $N$.
Given $\pi \in \Pi _{N}$, let $Pre\left( i,\pi \right) $ denote the set of
predecessors of $i$ in the order given by $\pi $, \textit{i.e.}, $Pre\left(
i,\pi \right) =\left\{ j\in N\mid \pi \left( j\right) <\pi \left( i\right)
\right\} .$ The \textit{Shapley value} (Shapley, 1953) is the most
well-known solution concept for $TU$ games. It is defined for each player as
the average of his contributions across orders of agents. Formally, for each 
$i\in N$, 
\[
Sh_{i}\left( N,v\right) =\frac{1}{n!}\sum_{\pi \in \Pi _{N}}\left[ v\left(
Pre\left( i,\pi \right) \cup \left\{ i\right\} \right) -v\left( Pre\left(
i,\pi \right) \right) \right] . 
\]

%\bigskip

We say that a decomposable index $I^d$ 
%$d:$ $N\times M\times \mathbb{N}_{+}^{N}\rightarrow \mathbb{R}% _{+}$ 
is \textbf{Shapley induced} if for all $j\in M$ and $x\in \mathbb{N}_{+}^{N}$
there exists a cooperative game $\left( N,v^{j,x}\right) $ such that for all 
$S\subset N$, 
\begin{equation}
v^{j,x}\left( S\right) =v^{j,\left( x_{i}\right) _{i\in S\cap L^{j}\left(
N,\left\{ j\right\} ,x\right) }}\left( S\cap L^{j}\left( N,\left\{ j\right\}
,x\right) \right),  \label{eq Sh ind}
\end{equation}%
and for all $i\in N$, $d\left( i,j,x\right) =Sh_{i}\left( N,v^{j,x}\right) $.

\bigskip Pro-rata and Shapley belong to this family of indexes, whereas
user-centric does not. In the case of the Shapley index the associated
cooperative game $\left( N,v^{j,x}\right) $ is just the game $\left(
N,v_{\left( N,\left\{ j\right\} ,x\right) }\right) $ introduced by Berganti%
\~{n}os and Moreno-Ternero (2025a). In the case of the pro-rata index, the
associated cooperative game $\left( N,v^{j,x}\right) $ is given by $%
v^{j,x}\left( S\right) =\sum\limits_{i\in S}x_{i}.$

\begin{theorem}
\label{char NA+AdU+EIA}An index satisfies additivity, null artists, and
equal impact of artists if and only if it is a Shapley-induced decomposable
index.
\end{theorem}

We conclude this section studying the combination of our two basic axioms
(additivity and null artists) with two of the three \textquotedblleft equal
impact\textquotedblright\ axioms. That is, we obtain the sub-families of
decomposable indices that satisfy each pair of \textquotedblleft equal
impact\textquotedblright\ axioms. Equivalently, we explore the intersections
of the sub-families of decomposable indices characterized above.

\begin{theorem}
\label{char two impact axioms} The following statements hold:

$\left( a\right) $ No decomposable index satisfies equal individual impact
of similar users and equal global impact of users.

$\left( b\right) $ A decomposable index satisfies equal individual impact of
similar users and equal impact of artists if an only if for all $i\in N$
there exists a function $f_{i}:\mathbb{N}_{+}\rightarrow \mathbb{R}_{+}$
such that for each $(i,j,x)\in N\times M\times\mathbb{N}_{+}^{N}$, $d\left(
i,j,x\right) =f_{i}\left( x_{i}\right)$.

$\left( c\right) $ A decomposable index satisfies equal global impact of
users and equal impact of artists if an only if it is the Shapley index.
\end{theorem}

Several lessons can be obtained from Theorem \ref{char two impact axioms}.
For instance, that no decomposable index satisfies the three equal impact
axioms. Also, that no index satisfies additivity, null artists, equal
individual impact of similar users and equal global impact of users. It was
already known (as a direct consequence of Proposition 1) that no index
satisfies additivity, pairwise homogeneity, equal individual impact of
similar users and equal global impact of users. Theorem \ref{char two impact
axioms} implies that this incompatibility remains when pairwise homogeneity
is weakened to null artists. Finally, another consequence of Theorem \ref%
{char two impact axioms} is that the intersection between the family of
aggregate-invariant decomposable indices and the family of Shapley-induced
decomposable indices is precisely the Shapley index. Thus, as a side effect, we have obtained
another axiomatic characterization of the Shapley index. 

The second statement of Theorem \ref{char two impact axioms} characterizes a
new (sub)family of decomposable indices. This family constitutes a natural
generalization of the pro-rata index, which arises when $f$ is precisely the
identity function. Any other function satisfying the condition $f(0)=0$
would give rise to another member within the family. In particular, the
(non-continuous) function that yields zero for any value below a given
threshold $\tau$, becoming the identity afterwards. That would give rise to
an index discussed below, which only considers the artists with
streamings above a given threshold $\tau$ for each user.

\section{Connections to the literature}

We now discuss the relationship between the new families introduced in this
paper and other existing families in the literature.

As mentioned above, Berganti\~{n}os and Moreno-Ternero (2025a) introduced
the family of weighted indices, which are included within the family of
decomposable indices introduced here. But, obviously, not the other way
around, as there exist weighted indices that do not depend only on artists.
Besides, there exist weighted indices that are not aggregate invariant.
Finally, each weighted index is Shapley induced, where for each $\left(
N,M,t\right) \in \mathcal{P}$, each $j\in \mathbb{M}$, each $x\in \mathbb{Z}%
_{+}^{n},$ and each $S\subset N,$ 
\[
v^{j,x}\left( S\right) =\sum_{i\in S}w\left( j,x\right) x_{i}. 
\]

%\bigskip

Berganti\~{n}os and Moreno-Ternero (2025a) also introduced the family of
probabilistic indices. To wit, a probability system is a function $\rho:%
\mathbb{M}\times \mathbb{Z}_{+}^{n}\rightarrow \mathbb{R}^{n}$ such that for
each $j\in \mathbb{M}$ and each $x\in \mathbb{Z}_{+}^{n}, $ $0\leq
\rho_{i}\left( j,x\right) \leq 1$, $\rho_{i}\left( j,x\right) =0 $ when $%
x_{i}=0,$ and $\sum\limits_{i=1}^{n}\rho_{i}\left( j,x\right) =1.$ For each
probability system $\rho$, the \textbf{probabilistic index }$I^{\rho }$ is
defined as follows. For each $\left( N,M,t\right) \in \mathcal{P}$ and each $%
i\in N,$ 
\[
I_{i}^{\rho}\left( N,M,t\right) =\sum_{j\in M}\rho_{i}\left( j,t_{.j}\right)
. 
\]

Berganti\~{n}os and Moreno-Ternero (2025a) show that probabilistic indices
satisfy additivity and reasonable lower bound (which is stronger than null
artists). This implies that each probabilistic index is a decomposable
index. But not the other way around as there exist probabilistic indices
that do not depend only on artists. Besides, there exist probabilistic
indices that are not aggregate invariant. Finally, each probabilistic index
is Shapley induced where for each $\left( N,M,t\right) \in \mathcal{P}$,
each $j\in \mathbb{M}$, each $x\in \mathbb{Z}_{+}^{n},$ and each $S\subset
N, $ 
\[
v^{j,x}\left( S\right) =\sum_{i\in S}\rho _{i}\left( j,x\right) . 
\]

%\bigskip

Gon\c{c}alves-Dosantos et al. (2025b) characterize two families of indices.
The first family is obtained by considering specific linear combinations of
the so-called equal-division index (which assigns the same importance to
each artist) 
%. Formally, for each problem $\left( N,M,t\right) $ and each artist $i\in N,$ $E_{i}\left( N,M,t\right) =\frac{m}{n}.$
and the user-centric index. Formally, for each $\beta \in \left[ 0,\frac{n}{%
n-1}\right] ,$ $I^{1,\beta }$ is defined so that for each $\left(
N,M,t\right) \in \mathcal{P}$ and each $i\in N,$ 
\[
I_{i}^{1,\beta }\left( N,M,t\right) =\beta E_{i}\left( N,M,t\right) +\left(
1-\beta \right) U_{i}\left( N,M,t\right) .
\]

Notice that the equal-division index $E$ does not satisfy null artists.
Consequently, only the user-centric index (which corresponds to the case $%
\beta =0)$ within this family is a decomposable index.

The second family is obtained by considering specific linear combinations of
the equal-division index and the reward associated to the pro-rata index.
Formally, for each $\beta \in \left[ 0,\frac{n}{n-1}\right] ,$ $I^{2,\beta }$
is defined so that for each $\left( N,M,t\right) \in \mathcal{P}$ and each $%
i\in N,$ 
\[
I_{i}^{2,\beta }\left( N,M,t\right) =\beta E_{i}\left( N,M,t\right) +\left(
1-\beta \right) R_{i}^{P}\left( N,M,t\right) .
\]

As $E$ does not satisfy null artists and $R^{P}$ does not satisfy additivity
(Berganti\~{n}os and Moreno-Ternero, 2025a) no index $I^{2,\beta }$ is a
decomposable index.

\section{The current debate in the music industry} 

Currently, there is an ongoing debate in the music industry about how artists should be
rewarded. As part of that debate, some big platforms (such as Spotify and Deezer) have decided to update their reward systems. We now discuss how the family of rules considered in this
paper could help the music industry to implement a fair reward system for artists. 

\subsection{Spotify's new royalties model}

In late 2023, Spotify announced that it is planning to implement a new
royalties model to \textquotedblleft drive more money to more popular
artists, record labels and distributors, while clamping down on streaming
fraud\textquotedblright .\footnote{%
See, for instance,
https://www.billboard.com/business/streaming/spotify-new-royalties-model-explained-how-work-1235501887/\#%
} 

One of the pillars of Spotify's new model is that fraudulent streams
will be charged. This can trivially be applied to any
reward system we could think of. Another pillar is that \textquotedblleft non-music noise
tracks\textquotedblright\ should be qualified in a different way for
royalties. In the parlance of our paper, this simply means that the matrix $t$ should be computed in a different way. It does not say nothing about the reward system itself. 

More interestingly, another pillar is that tracks that receive less than 1000 streams will not qualify for royalties.\footnote{In our model, we consider artists instead of tracks. But we could consider tracks instead of artists and all of our results would remain valid.}. Thus, in the parlance of our paper, Spotify is using the following index: 
\[
Sp_{i}\left( N,M,t\right) =\left\{ 
\begin{tabular}{ll}
$T_{i}$ & if $T_{i}\geq 1000$ \\ 
0 & if $T_{i}<1000.$%
\end{tabular}%
\right. 
\]%
It is trivial to check that this index satisfies null artists but fails to satisfy
additivity. Thus, it does not belong to any of the families considered above. %Now, the idea underlying this pillar is that some artists are irrelevant (those with less that 1000 streams) and, consequently, they will not be rewarded. 
Now, there exist reasons to avoid this index. Suppose, for instance, that artist $i$ received $800$ streams, half from user $1$ and half from user $2$. Besides, assume both users only streamed artist $i$. Then, artist $i$ might leave the platform, which would prompt users $1$ and $2$ to leave the platform too. This motivates an alternative way to accommodate the idea behind the above pillar, which would be compatible with the families we characterize in this paper. To wit, instead of considering when an artist is not relevant for the
whole set of users (as Spotify did), we consider when an artist is not
relevant for a single user. That is, we are suggesting the separable index that only considers artists with streamings above a given threshold $\tau$, for each user. Formally,
let $I^{d}$ be the index arising from 
\[
d\left( i,j,x\right) =\left\{ 
\begin{tabular}{ll}
$x_{i}$ & if $x_{i}\geq \tau $ \\ 
0 & if $x_{i}<\tau .$%
\end{tabular}%
\right. 
\]

%With this index artist $i$ will be paid when $\tau <400.$ \bigskip 
\subsection{Deezer's artist-centric method}
%Deezer has been very in the last years.\footnote{See, for instance,
%\par
%https://newsroom-deezer.com/2019/09/deezer-launches-initiative-to-make-the-music-industry-fairer-for-artists-2/ and https://www.universalmusic.com/universal-music-group-and-deezer-to-launch-the-first-comprehensive-artist-centric-music-streaming-model/.} 
%In 2019, Deezer already launched an online initiative to explain the benefits of moving from the pro-rata to the user-centric method. 
In 2023, Deezer launched
the so-called artist-centric method \textquotedblleft to create fairer
incentives and shift more revenue towards the creators who make the music
users love\textquotedblright .

It is based on four pillars. The first pillar is \textquotedblleft distinguishing between music and noise\textquotedblright . In the parlance of our paper, this simply means that the matrix $t$ should be computed in a different way (to avoid noise) but it
says nothing about the reward system itself. The second pillar says
\textquotedblleft artists who amass over 1000 streams per month from at
least 500 listeners receive a double boost\textquotedblright . Again, in the parlance of our paper, this also means
that the matrix $t$ should be computed in a different way, but it says
nothing about the reward system either. The third pillar aims to \textquotedblleft
prioritize active streams over algorithmic manipulation\textquotedblright .
Thus, same conclusion as before. 

Now, the fourth pillar aims to \textquotedblleft implement a user cap to prevent
system abuse upon limiting streams per user to 1000 and incorporating a
robust fraud detection system\textquotedblright . The second part (referring to the fraud
detection) is again related to the way matrix $t$ is defined (the fraudulent streams are not
qualified for royalties). As for the first part, Deezer does not say explicitly how to
implement the user cap. But we believe that the most natural way to do so is
to consider a proportional cap. Namely, assume that user $j$ has more than $1000$
streams. Then, for each artist $i$, we consider the corresponding truncated streams: $t_{ij}^{\prime }=\frac{t_{ij}}{T^{j}}1000$. 

The family of aggregate-invariant decomposable indices is aligned with this
pillar because none of those indices are affected by truncation. Namely, for
each $\left( N,M,t\right)\in\mathcal{P} $, $I^{d}\left( N,M,t\right) =I^{d}\left(
N,M,t^{\prime }\right) .$ Thus, the indices already prevent this abuse and one does not need to limit streams. 

On the other hand, some indices within the family of user-independent decomposable indices do not
prevent this abuse (for instance, the pro-rata index) but some others do prevent it. For
instance, consider the index induced by 
\[
d\left( i,j,x\right) =\left\{ 
\begin{tabular}{cc}
$x_{i}$ & if $\sum\limits_{i^{\prime }\in N}x_{i^{\prime }}<1000$ \\ 
$\frac{x_{i}}{\sum\limits_{i^{\prime }\in N}x_{i^{\prime }}}1000$ & if $%
\sum\limits_{i^{\prime }\in N}x_{i^{\prime }}>1000.$%
\end{tabular}%
\right. 
\]

In this index, the \textquotedblleft importance\textquotedblright\ of users
with less that 1000 streams is just the number of streams the artists had, whereas the
importance of users with more that 1000 streams is just 1000, thus preventing the
abuse. 

Likewise, within the family of Shapley induced decomposable indices, some indices do not
prevent the abuse (for instance, the pro-rata index) whereas some others do prevent it (for
instance, the Shapley index). 

To summarize, we can safely argue that our family of decomposable indices can accommodate
Deezer's pillars in several ways.\footnote{Moreau et al. (2024) also consider a reward system that accommodates Deezer's pillars, and they compare it empirically with the pro-rata and user-centric systems.}

\section{Discussion}

In this paper, we have focussed on the allocation problem that arises in
music platforms to reward artists from the payments users make to access the
platform. This issue is receiving increasing attention lately, to the extent
of becoming a major concern for platforms and users alike (e.g., Jensen,
2024). For instance, as discussed in the previous section, Spotify and Deezer launched new methods aimed at implementing fairer rewards for artists. Apart from revealing
the importance platforms devote to this issue, these moves also suggest the
possibility that platforms may compete on the basis of the reward systems
they offer.\footnote{Etro (2023) has recently analyzed a different form of competition among platforms. Tirole (2023) surveys the literature on the more general issue of competition in the digital era.} In any case, it seems to be clear that more study on reward
systems is needed, and we modestly believe our work contributes to that
goal.

The starting point for our axiomatic analysis is the characterization of the
broad family of decomposable indices by means of two basic axioms
(additivity and null teams). We have also shown that each one of a trio of
axioms formalizing homogeneity of streams, users and artists, respectively,
narrows the family of decomposable indices in meaningful ways. More
precisely, we characterize the sub-families of user-independent decomposable
indices, aggregate-invariant decomposable indices and Shapley-induced
decomposable indices. The intersection of the first two sub-families is
empty, whereas the intersection of the last two is precisely the Shapley
index. Finally, the intersection of the first and third sub-families gives
rise to an interesting generalization of the classical pro-rata index,
allowing for indices implementing thresholds on the number of streamings artists get. We
have also explored how these new families connect to existing families in
the literature (such as compromises between the pro-rata and the
user-centric indices, or generalizations of the latter) and even to the new methods recently launched by Spotify and Deezer. Our results thus
provide normative foundations for new methods to compensate artists in music
streaming platforms.

\bigskip 

\bigskip 

\bigskip 

\section{Appendix}

\subsection{Proof of Theorem \protect\ref{char NA+AdU}}

%\begin{proof}
We first prove that each decomposable index satisfies additivity and null
artists.

%For each $d$, $I^{d}$ satisfies additivity. 
Let $\left( N,M^{1},t^{1}\right)\in \mathcal{P}$, $\left(
N,M^{2},t^{2}\right)\in \mathcal{P} $, and $\left( N,M,t\right)\in \mathcal{P%
} $ be as in the definition of additivity. For each $i\in N$, 
\begin{eqnarray*}
I_{i}^{d}\left( N,M^{1},t^{1}\right) +I_{i}^{d}\left( N,M^{2},t^{2}\right)
&=&\sum_{j\in M^{1}}d\left( i,j,t^{1}._{j}\right) +\sum_{j\in M^{2}}d\left(
i,j,t^{2}._{j}\right) \\
&=&\sum_{j\in M}d\left( i,j,t._{j}\right) =I_{i}^{d}\left( N,M,t\right) .
\end{eqnarray*}

%For each $d$, $I^{d}$ satisfies null artists. 
Let $\left( N,M,t\right)\in \mathcal{P} $, and $i\in N$ be such that $T_{i}=0
$. As $t_{ij}=0$ for all $j\in M$, it follows by condition 1 of the
definition of decomposable index that $d\left( i,j,t._{j}\right) =0$. Thus, 
\[
I_{i}^{d}\left( N,M,t\right) =\sum_{j\in M}d\left( i,j,t._{j}\right) =0. 
\]

We now prove that if an index $I$ satisfies additivity and null artists,
then $I$ is a decomposable index.

For each $i\in N$, each $j\in M$, and each $x\in $ $\mathbb{N}_{+}^{N}$, we
define $d\left( i,j,x\right) =I_{i}\left( N,\left\{ j\right\} ,x\right) .$
Assume that $x_{i}=0.$ As $I$ satisfies null artists, $I_{i}\left( N,\left\{
j\right\} ,x\right) =0$ and hence $d\left( i,j,x\right) =0.$

Let $\left( N,M,t\right)\in \mathcal{P}$. By additivity, for each $i\in N,$ 
\[
I_{i}\left( N,M,t\right) =\sum_{j\in M}I_{i}\left( N,\left\{ j\right\}
,t_{.j}\right) =\sum_{j\in M}d\left( i,j,t_{.j}\right) . 
\]

Thus, $I=I^{d}$, as desired. This concludes the proof of statement (a). 
%\end{proof}

\bigskip

Note that both axioms at statement (a) are essential. The equal-division
index is an example of a non-decomposable index that satisfies additivity
but fails null artists. Likewise, the index in which the importance of each
artist is the square of the streams is an example of a non-decomposable
index that satisfies null artists but fails additivity.

%\begin{remark} \label{ind NA+AD}The axioms used in Theorem \ref{char NA+AdU} are independent.
%$\left( a\right) $ The equal-division index $E$ defined in Remark \ref{remark equal-division} satisfies additivity but fails null artists.
%Let $I^{1}$ be the index in which the importance of each artist is the square of the streams. Namely, for each $\left( N,M,t\right) $ and each $i\in N,$ 
%\[ I_{i}^{1}\left( N,M,t\right) =\left( T_{i}\right) ^{2}.\]
%$I^{1}$ satisfies null artists but fails additivity.
%\end{remark}
\bigskip

%\begin{proof}
Let now $I$ be a user-independent decomposable index. By statement (a), 
%Theorem \ref{char NA+AdU}, 
$I$ satisfies \textit{additivity} and \textit{null artists}. As for \textit{%
equal individual impact of similar users}, let $\left( N,M,t\right) \in 
\mathcal{P}$, $i\in N,$ and $j,j^{\prime }\in M$ be such that $%
t_{ij}=t_{ij^{\prime }}$. Notice that, 
\[
I_{i}\left( N,M,t\right) =\sum_{j^{\ast }\in M}d\left( i,j^{\ast
},t._{j^{\ast }}\right) =I_{i}\left( N,M\backslash \left\{ j\right\}
,t^{-j}\right) +d\left( i,j,t._{j}\right), 
\]
and 
\[
I_{i}\left( N,M,t\right) =\sum_{j^{\ast }\in M}d\left( i,j^{\ast
},t._{j^{\ast }}\right) =I_{i}\left( N,M\backslash \left\{ j^{\prime
}\right\} ,t^{-j^{\prime }}\right) +d\left( i,j^{\prime },t._{j^{\prime
}}\right) . 
\]

As $I^d$ is user independent, $d\left( i,j,t._{j}\right) =d\left(
i,j^{\prime },t._{j^{\prime }}\right)$. Then, 
\[
I_{i}\left( N,M\backslash \left\{ j\right\} ,t^{-j}\right) =I_{i}\left(
N,M\backslash \left\{ j^{\prime }\right\} ,t^{-j^{\prime }}\right). 
\]
%and hence $I$ satisfies \textit{equal individual impact of similar users}.

We now prove that if an index $I$ satisfies \textit{additivity}, \textit{%
null artists}, and \textit{equal individual impact of similar users} then $I$
is a user-independent decomposable index. 
%Let $I$ be an index satisfying the three axioms. 
By statement (a), %Theorem \ref{char NA+AdU} 
$I$ is a decomposable index. %Namely $I=I^{d}$ for some function $d$. 
We now prove that it is also user independent. Let $\left( i,j,x\right) $
and $\left( i,j^{\prime },x^{\prime }\right) $ be such that $%
x_{i}=x_{i}^{\prime }.$ Let $\left( N,M,t\right)\in\mathcal{P} $ be such
that $i\in N,$ $M=\left\{ j,j^{\prime }\right\} ,$ and for all $i^{\prime
}\in N,$ $t_{i^{\prime }j}=x_{i^{\prime }}$ and $t_{i^{\prime }j^{\prime
}}=x_{i^{\prime }}^{\prime }.$ As $I$ is a decomposable index, $I_{i}\left(
N,M\backslash \left\{ j\right\} ,t^{-j}\right) =d\left( i,j^{\prime
},x^{\prime }\right)$, and $I_{i}\left( N,M\backslash \left\{ j^{\prime
}\right\} ,t^{-j^{\prime }}\right)= d\left( i,j,x\right)$. Finally, as $I$
satisfies \textit{equal individual impact of similar users}, $d\left(
i,j,x\right) =d\left( i,j^{\prime },x^{\prime }\right)$. This concludes the
proof of statement (b). %\end{proof}

\bigskip

Note that the three axioms at statement (b) are essential. The
equal-division index is an example of a non-decomposable index that
satisfies all axioms but \textit{null artists}. Likewise, user-centric index
satisfies all axioms but \textit{equal individual impact of similar users}.
Finally, the following index satisfies all axioms but \textit{additivity}.
For each $\left( N,M,t\right)\in\mathcal{P} $ and each $i\in N,$ 
\[
I_{i}^{2}\left( N,M,t\right) =\sum_{j=1}^{m}\frac{t_{ij}+T_{i}}{
T^{j}+\sum\limits_{i^{\prime }\in N}T_{i^{\prime }}}. 
\]

%\begin{remark} \label{ind NA+AD+IISU}The axioms used in Theorem \ref{char NA+AdU+EIISU} are independent.

%The equal-division index satisfies .

%Let $I^{2}$ be defined as follows. For each $\left( N,M,t\right) $ and each $i\in N,$ 
%\[I_{i}^{2}\left( N,M,t\right) =\sum_{j=1}^{m}\frac{t_{ij}+T_{i}}{T^{j}+\sum\limits_{i^{\prime }\in N}T_{i^{\prime }}}.\]

%$I^{2}$ satisfies all axioms but \textit{additivity}.

%\end{remark}

%\bigskip

%begin{proof}
Let now $I$ be an aggregate-invariant decomposable index. By statement (a), $%
I$ satisfies \textit{additivity} and \textit{null artists}. As for \textit{%
equal global impact of users}, let $\left( N,M,t\right) \in \mathcal{P}$ and 
$j,j^{\prime }\in M$. Notice that, 
\[
\sum\limits_{i\in N}I_{i}\left( N,M,t\right) =\sum\limits_{i\in
N}I_{i}\left( N,M\backslash \left\{ j\right\} ,t^{-j}\right)
+\sum\limits_{i\in N}d\left( i,j,t._{j}\right) 
\]
and 
\[
\sum\limits_{i\in N}I_{i}\left( N,M,t\right) =\sum\limits_{i\in
N}I_{i}\left( N,M\backslash \left\{ j^{\prime }\right\} ,t^{-j^{\prime
}}\right) +\sum\limits_{i\in N}d\left( i,j^{\prime },t._{j^{\prime
}}\right). 
\]

As $d$ is aggregate invariant, $\sum\limits_{i\in N}d\left(
i,j,t._{j}\right) =\sum\limits_{i\in N}d\left( i,j^{\prime },t._{j^{\prime
}}\right) .$ Then, 
\[
\sum\limits_{i\in N}I_{i}\left( N,M\backslash \left\{ j\right\}
,t^{-j}\right) =\sum\limits_{i\in N}I_{i}\left( N,M\backslash \left\{
j^{\prime }\right\} ,t^{-j^{\prime }}\right). 
\]
%and hence $I$ satisfies equal global impact of users.

We now prove that if an index $I$ satisfies \textit{additivity}, \textit{%
null artists}, and \textit{equal global impact of users} then $I$ is an
aggregate-invariant decomposable index. Let $I$ be an index satisfying the
three axioms. By statement (a), $I$ is a decomposable index. We now prove
that it is also aggregate invariant. Let $j,j^{\prime }\in M$ and $%
x,x^{\prime }\in \mathbb{N}_{+}^{N}.$ Let $\left( N,M,t\right)\in\mathcal{P} 
$ be such that $M=\left\{ j,j^{\prime }\right\} ,$ and for all $i\in N,$ $%
t_{ij}=x_{i}$ and $t_{ij^{\prime }}=x_{i}^{\prime }.$ As $I$ is a
decomposable index, 
\[
\sum\limits_{i\in N}I_{i}\left( N,M\backslash \left\{ j\right\}
,t^{-j}\right) =\sum\limits_{i\in N}d\left( i,j^{\prime },x^{\prime
}\right), 
\]
and 
\[
\sum\limits_{i\in N}I_{i}\left( N,M\backslash \left\{ j^{\prime }\right\}
,t^{-j^{\prime }}\right) =\sum\limits_{i\in N}d\left( i,j,x\right) . 
\]

As $I$ satisfies \textit{equal global impact of users}, $\sum\limits_{i\in
N}d\left( i,j,x\right) =\sum\limits_{i\in N}d\left( i,j^{\prime },x^{\prime
}\right)$. This concludes the proof of statement (c). % \end{proof}

\bigskip

Note that the three axioms at statement (c) are essential. The
equal-division index is an example of a non-decomposable index that
satisfies all axioms but \textit{null artists}. Likewise, the pro-rata index
satisfies all axioms but \textit{equal global impact of users}. Finally, the
index assigning all the revenues to the artist with more streams satisfies
all axioms but \textit{additivity}. \hfill $\Box$

%\bigskip

%\begin{remark}\label{ind NA+AD+EGIU}The axioms used in Theorem \ref{char NA+AdU+EIISU} are independent.

%The equal-division index satisfies all axioms but \textit{null artists}.

%Let $I^{3}$ be the index assigning all the revenues to the artist with more streams. For each $\left( N,M,t\right) $ let $H$ be the set of artists with highest number of streams. Then, 
%\[
%I_{i}^{3}\left( N,M,t\right) =\left\{ 
%\begin{tabular}{ll}
%$\frac{m}{\left\vert H\right\vert }$ & if $i\in H$ \\ 
%0 & otherwise$.$%
%\end{tabular}%
%\right. 
%\]

%$I^{3}$ satisfies all axioms but \textit{additivity}.

%The pro-rata index satisfies all axioms but equal global impact of users.
%\end{remark}
%\bigskip

\subsection{Proof of Theorem \protect\ref{char NA+AdU+EIA}}

%\begin{proof}
Let $I$ be a Shapley-induced decomposable index. By Theorem \ref{char NA+AdU}%
, $I$ satisfies \textit{additivity} and \textit{null artists}. As for 
\textit{equal impact of artists}, let $\left( N,M,t\right) \in \mathcal{P}$
and $i,i^{\prime }\in N$. Notice that, 
\[
I_{i}\left( N,M,t\right) -I_{i}\left( N\backslash \left\{ i^{\prime
}\right\} ,M,t_{-i^{\prime }}\right) =\sum_{j\in M}\left( d\left(
i,j,t._{j}\right) -d\left( i,j,\left( t_{-i^{\prime }}\right) ._{j}\right)
\right), 
\]
and 
\[
I_{i^{\prime }}\left( N,M,t\right) -I_{i^{\prime }}\left( N\backslash
,M,t_{-i}\right) =\sum_{j\in M}\left( d\left( i^{\prime },j,t._{j}\right)
-d\left( i^{\prime },j,\left( t_{-i}\right) ._{j}\right) \right) . 
\]

Then, it is enough to prove that, for each $j\in M,$ 
\begin{equation}
d\left( i,j,t._{j}\right) -d\left( i,j,\left( t_{-i^{\prime }}\right)
._{j}\right) =d\left( i^{\prime },j,t._{j}\right) -d\left( i^{\prime
},j,\left( t_{-i}\right) ._{j}\right)  \label{eq imp art}
\end{equation}

As $I$ is Shapley induced, we have that 
\begin{eqnarray*}
d\left( i,j,t._{j}\right) &=&Sh_{i}\left( N,v^{j,t._{j}}\right) \\
d\left( i,j,\left( t_{-i^{\prime }}\right) ._{j}\right) &=&Sh_{i}\left(
N\backslash \left\{ i^{\prime }\right\} ,v^{j,\left( t_{-i^{\prime
}}\right), ._{j}}\right), \\
d\left( i^{\prime },j,t._{j}\right) &=&Sh_{i^{\prime }}\left(
N,v^{j,t._{j}}\right), \\
d\left( i^{\prime },j,\left( t_{-i}\right) ._{j}\right) &=&Sh_{i}\left(
N\backslash \left\{ i\right\} ,v^{j,\left( t_{-i}\right) ._{j}}\right) .
\end{eqnarray*}
By $\left( \ref{eq Sh ind}\right) ,$ for each $i^{\prime }\in N$ and each $%
S\subset N\backslash \left\{ i^{\prime }\right\} ,$ 
\begin{eqnarray*}
v^{j,t._{j}}\left( S\right) &=&v^{j,\left( t_{i^{\ast }j}\right) _{i^{\ast
}\in S\cap L^{j}\left( N,\left\{ j\right\} ,t._{j}\right) }}\left( S\cap
L^{j}\left( N,\left\{ j\right\} ,t._{j}\right) \right) \\
&=&v^{j,\left( t_{i^{\ast }j}\right) _{i^{\ast }\in S\cap L^{j}\left(
N\backslash \left\{ i^{\prime }\right\} ,\left\{ j\right\} ,\left(
t_{-i}\right) ._{j}\right) }}\left( S\cap L^{j}\left( N\backslash \left\{
i^{\prime }\right\} ,\left\{ j\right\} ,\left( t_{-i^{\prime }}\right)
._{j}\right) \right) \\
&=&v^{j,\left( t_{-i^{\prime }}\right) ._{j}}\left( S\right) .
\end{eqnarray*}
Thus, equation $\left( \ref{eq imp art}\right) $ can be rewritten as 
\[
Sh_{i}\left( N,v^{j,t._{j}}\right) -Sh_{i}\left( N\backslash \left\{
i^{\prime }\right\} ,v^{j,t._{j}}\right) =Sh_{i^{\prime }}\left(
N,v^{j,t._{j}}\right) -Sh_{i^{\prime }}\left( N\backslash \left\{ i\right\}
,v^{j,t._{j}}\right), 
\]
which is precisely the property of balanced contributions that characterizes
the Shapley value (e.g., Myerson, 1980).

We now prove that if an index $I$ satisfies \textit{additivity}, \textit{%
null artists}, and \textit{equal impact of artists} then $I$ is a
Shapley-induced decomposable index. By Theorem \ref{char NA+AdU}, $I$ is a
decomposable index. We now prove that it is Shapley induced.

Let $j\in M$ and $x\in \mathbb{N}_{+}^{N}.$ We proceed by induction on $%
\left\vert N\right\vert .$ If $N=\left\{ i\right\} ,$ we define $%
v^{j,x}\left( i\right) =I_{i}\left( N,\left\{ j\right\} ,x\right) .$ Then, 
\[
d\left( i,j,x\right) =I_{i}\left( N,\left\{ j\right\} ,x\right)
=Sh_{i}\left( N,v^{j,x}\right) . 
\]

Assume that if $\left\vert N\right\vert \leq p,$ $d\left( i,j,x\right)
=Sh_{i}\left( N,v^{j,x}\right) $ where $\left( N,v^{j,x}\right) $ satisfies $%
\left( \ref{eq Sh ind}\right) $. We prove it for $\left\vert N\right\vert
=p+1.$

As $I$ satisfies equal impact of artists, we have that, for each pair $%
i,i^{\prime }\in N$, 
\[
I_{i}\left( N,\left\{ j\right\} ,x\right) -I_{i}\left( N\backslash \left\{
i^{\prime }\right\} ,\left\{ j\right\} ,x_{-i^{\prime }}\right)
=I_{i^{\prime }}\left( N,\left\{ j\right\} ,x\right) -I_{i^{\prime }}\left(
N\backslash \left\{ i\right\} ,\left\{ j\right\} ,x_{-i}\right) . 
\]
Thus, 
\[
I_{i}\left( N,\left\{ j\right\} ,x\right) -I_{i^{\prime }}\left( N,\left\{
j\right\} ,x\right) =I_{i}\left( N\backslash \left\{ i^{\prime }\right\}
,\left\{ j\right\} ,x_{-i^{\prime }}\right) -I_{i^{\prime }}\left(
N\backslash \left\{ i\right\} ,\left\{ j\right\} ,x_{-i}\right) . 
\]

Hence, for each $i\in N$, 
\[
nI_{i}\left( N,\left\{ j\right\} ,x\right) -\sum\limits_{i^{\prime }\in
N}I_{i^{\prime }}\left( N,\left\{ j\right\} ,x\right)
=\sum\limits_{i^{\prime }\in N}\left[ I_{i}\left( N\backslash \left\{
i^{\prime }\right\} ,\left\{ j\right\} ,x_{-i^{\prime }}\right)
-I_{i^{\prime }}\left( N\backslash \left\{ i\right\} ,\left\{ j\right\}
,x_{-i}\right) \right] . 
\]
Or, equivalently, 
\begin{equation}
\left( n-1\right) I_{i}\left( N,\left\{ j\right\} ,x\right)
-\sum\limits_{i^{\prime }\in N\backslash \left\{ i\right\} }I_{i^{\prime
}}\left( N,\left\{ j\right\} ,x\right) = \sum\limits_{i^{\prime }\in N}\left[
I_{i}\left( N\backslash \left\{ i^{\prime }\right\} ,\left\{ j\right\}
,x_{-i^{\prime }}\right) -I_{i^{\prime }}\left( N\backslash \left\{
i\right\} ,\left\{ j\right\} ,x_{-i}\right) \right] .  \label{eq bal cont}
\end{equation}

By the induction hypothesis, for each pair $i,i^{\prime }\in N$, 
\[
I_{i}\left( N\backslash \left\{ i^{\prime }\right\} ,\left\{ j\right\}
,x_{-i^{\prime }}\right) = Sh_{i}\left( N\backslash \left\{ i^{\prime
}\right\} ,v^{j,x_{-i^{\prime }}}\right), 
\]
and 
\[
I_{i^{\prime }}\left( N\backslash \left\{ i\right\} ,\left\{ j\right\}
,x_{-i}\right) = Sh_{i}\left( N\backslash \left\{ i\right\}
,v^{j,x_{-i}}\right) . 
\]
%By $\left( \ref{eq bal cont}\right) $ 
Thus, we obtain a system of $n$ independent equations (each $i\in N$ yields
one equation) and $n$ variables $\left( I_{i}\left( N,\left\{ j\right\}
,x\right)\right)_{i\in N}$, with a unique solution. Therefore, $I\left(
N,\left\{ j\right\} ,x\right) $ is uniquely determined. We prove that such a
unique solution is given by a Shapley-induced decomposable index $I^d$.

For each $S\subsetneq N$, we define 
\[
v^{j,x}\left( S\right) =v^{j,\left( x_{i}\right) _{i\in S\cap L^{j}\left(
N,\left\{ j\right\} ,x\right) }}\left( S\cap L^{j}\left( N,\left\{ j\right\}
,x\right) \right). 
\]
If $L^{j}\left( N,\left\{ j\right\} ,x\right) \neq N$, we define 
\[
v^{j,x}\left( N\right) =v^{j,\left( x_{i}\right) _{i\in L^{j}\left(
N,\left\{ j\right\} ,x\right) }}\left( L^{j}\left( N,\left\{ j\right\}
,x\right) \right). 
\]
Finally, if $L^{j}\left( N,\left\{ j\right\} ,x\right) =N$, we define 
\[
v^{j,x}\left( N\right) =\sum\limits_{i\in N}I_{i}\left( N,\left\{ j\right\}
,x\right). 
\]

It follows by the induction hypothesis that $v^{j,x}$, so defined, satisfies 
$\left( \ref{eq Sh ind}\right)$. 
%If $S\subsetneq N$ or $S=N\neq L^{j}\left( N,\left\{ j\right\} ,x\right) ,$ by induction hypothesis, $v^{j,x}$ satisfies $\left( \ref{eq Sh ind}\right) .$ If $S=N=L^{j}\left( N,\left\{ j\right\} ,x\right) ,$ then $\left( \ref{eq Sh ind}\right) $ holds trivially.

Let $i^{\prime }\in N$. By the induction hypothesis, for each $S\subset
N\backslash \left\{ i^{\prime }\right\}$, 
\[
v^{j,x_{-i^{\prime }}}\left( S\right) =v^{j,\left( x_{i}\right) _{i\in S\cap
L^{j}\left( N,\left\{ j\right\} ,x\right) }}\left( S\cap L^{j}\left(
N,\left\{ j\right\} ,x\right) \right) =v^{j,x}\left( S\right). 
\]

Then, for each $i^{\prime }\in N,$ $\left( N\backslash \left\{ i^{\prime
}\right\} ,v^{j,x_{-i^{\prime }}}\right) =\left( N\backslash \left\{
i^{\prime }\right\} ,v^{j,x}\right) .$ Now, for each $i\in N$ equation $%
\left( \ref{eq bal cont}\right) $ can be rewritten as follows: 
\begin{equation}
\left( n-1\right) I_{i}\left( N,\left\{ j\right\} ,x\right)
-\sum\limits_{i^{\prime }\in N\backslash \left\{ i\right\} }I_{i^{\prime
}}\left( N,\left\{ j\right\} ,x\right) =\sum\limits_{i^{\prime }\in N}\left[
Sh_{i}\left( N\backslash \left\{ i^{\prime }\right\} ,v^{j,x}\right)
-Sh_{i^{\prime }}\left( N\backslash \left\{ i\right\} ,v^{j,x}\right) \right]
.  \label{eq bal cont 2}
\end{equation}

Now, as mentioned above, the Shapley value satisfies balanced contributions
(e.g., Myerson, 1980), which says that for each cooperative game $\left(
N,v\right) $ and each pair $i,j\in N,$ 
\[
Sh_{i}\left( N,v\right) -Sh_{i}\left( N\backslash \left\{ j\right\}
,v\right) =Sh_{j}\left( N,v\right) -Sh_{j}\left( N\backslash \left\{
i\right\} ,v\right). 
\]
Consequently, we have that $I\left( N,\left\{ j\right\} ,x\right) =Sh\left(
N,v^{j,x}\right) $ satisfies $\left( \ref{eq bal cont 2}\right) .$ As $%
I\left( N,\left\{ j\right\} ,x\right) $ is uniquely determined, we deduce
that $I\left( N,\left\{ j\right\} ,x\right) =Sh\left( N,v^{j,x}\right)$.
Then, for all $i\in N,$ 
\[
d\left( i,j,x\right) =I_{i}\left( N,\left\{ j\right\} ,x\right)
=Sh_{i}\left( N,v^{j,x}\right), 
\]
as desired. \hfill $\Box$ %\end{proof}

\bigskip

Note that the three axioms in the statement are essential. The
equal-division index is an example of a non-decomposable index that
satisfies all axioms but \textit{null artists}. Likewise, the user-centric
index satisfies all axioms but \textit{equal impact of artists}. Finally,
the index assigning zero importance to null artists and one importance to
all the remaining artists satisfies all axioms but \textit{additivity}.

%\begin{remark} \label{ind NA+AD+EIA}The axioms used in Theorem \ref{char NA+AdU+EIA} are independent. The equal-division index satisfies all axioms but \textit{null artists}.
%Let $I^{4}$ be the index where \textit{null artists} have no importance and the rest of artists have the same importance. For each $\left( N,M,t\right) $ and each $i\in N,I_{i}^{4}\left( N,M,t\right) =1$ if $T_{i}>0$ and $I_{i}^{4}\left( N,M,t\right) =0$ if $T_{i}=0.$ $I^{4}$ satisfies all axioms but \textit{additivity}.
% The user-centric index satisfies all axioms but equal impact of artists.
%\end{remark}

\subsection{Proof of Theorem \protect\ref{char two impact axioms}}

% \begin{proof}
%$\left( a\right) $ Let $I$ be a decomposable index satisfying the two axioms.

We note first that statement (c) is obtained from Corollary 1 at Berganti%
\~{n}os and Moreno-Ternero (2025b). We thus concentrate on the other two
statements.

Assume, by contradiction, that $I^d$ is a decomposable index that satisfies 
\textit{equal individual impact of similar users} and \textit{equal global
impact of users}. That is, $I^d$ is a user-independent and
aggregate-invariant decomposable index.

Let $\left( N,M,t\right)\in\mathcal{P} $ and $j\in M$. For each $S\subset N$, let $1_{S}$
be the vector $\left( x_{i}\right) _{i\in N}$ such that $x_{i}=1$ whenever $%
i\in S$ and $x_{i}=0$ otherwise.

Let $i\in N$. Then, $d\left( i^{\prime },j,1_{i}\right) =0$ for all $%
i^{\prime }\neq i$. Hence, 
\[
\sum_{i^{\prime }\in N}d\left( i^{\prime },j,1_{i}\right) =d\left(
i,j,1_{i}\right) . 
\]

Now, as $I^d$ is user independent, it follows that $d\left( i,j,1_{i}\right)
=d\left( i,j,1_{N}\right) $ for all $i\in N.$

And as $I^d$ is aggregate invariant, it follows that 
\[
\sum_{i^{\prime }\in N}d\left( i^{\prime },j,1_{i}\right) =\sum_{i^{\prime
}\in N}d\left( i^{\prime },j,1_{i}\right). 
\]
Then, for all $i\in N,$ 
\[
\sum_{i^{\prime }\in N}d\left( i^{\prime },j,1_{N}\right) =\sum_{i^{\prime
}\in N}d\left( i^{\prime },j,1_{N}\right) =d\left( i,j,1_{i}\right) =d\left(
i,j,1_{N}\right) . 
\]
Thus, $d\left( i,j,1_{N}\right) =0$ for all $i\in N$ which represents a
contradiction. %contradicts the definition of index.
\bigskip

%$\left( b\right) $ 
Let now $I$ be a decomposable index induced by a function $f_{i}$ as in
statement (b). Then, it is straightforward to see that it satisfies the two
axioms in the statement (\textit{equal individual impact of similar users}
and \textit{equal impact of artists}). Conversely, let $I^d$ be a
decomposable index satisfying the two axioms. That is, $I^d$ is user
independent and Shapley induced. For each artist $i\in N$, let 
\[
f_{i}\left( x_{i}\right) =I^d_{i}\left( i,j,x_{i}\right)=d\left(
i,j,x_{i}\right) . 
\]
As $I^d$ is user independent, $d\left( i,j,x_{i}\right) =d\left( i,j^{\prime
},x_{i}\right) $ for all $j^{\prime }\neq j.$ Thus, $f_{i}$ is well defined
because it does not depend on $j$.

As $I^d$ is Shapley induced, $d\left( i,j,x_{i}\right) =Sh_{i}\left(
i,v^{j,x_{i}}\right) =v^{j,x_{i}}\left( i\right)$.

Let $i\in N$, $j\in M$ and $x\in \mathbb{N}_{+}^{N}$. As $I^d$ is user
independent, $d\left( i,j,x\right) =d\left( i,j,x_{i}1_{i}\right)$. And as $%
I^d$ is Shapley induced, %Theorem \ref{char NA+AdU+EIA}, 
\[
d\left( i,j,x_{i}1_{i}\right) =Sh_{i}\left( N,v^{j,x_{i}1_{i}}\right) . 
\]
As $L_{j}\left( N,j,x_{i}1_{i}\right) =\left\{ i\right\} $, equation $\left( %
\ref{eq Sh ind}\right) $ implies 
\[
v^{j,x_{i}1_{i}}\left( S\right) =\left\{ 
\begin{tabular}{cc}
0 & if $i\notin S$ \\ 
$v^{j,x_{i}}\left( i\right) $ & if $i\in S.$%
\end{tabular}%
\right. 
\]

Then, $Sh_{i}\left( N,v^{j,x_{i}1_{i}}\right) =v^{j,x_{i}}\left( i\right) .$
Combining the equalities obtained above we have that $d\left( i,j,x\right)
=f_{i}\left( x_{i}\right)$, as desired. %\bigskip

\hfill $\Box$ %\end{proof}

\bigskip

\bigskip

\end{document}